%% file: main.tex
\documentclass[10pt,twocolumn,letterpaper]{article}

\usepackage{cvpr}              % To produce the CAMERA-READY version
\input{preamble}

\definecolor{cvprblue}{rgb}{0.21,0.49,0.74}
\usepackage[accsupp]{axessibility}
\usepackage[pagebackref,breaklinks,colorlinks,allcolors=cvprblue]{hyperref}

\def\paperID{63} % *** Enter the Paper ID here
\def\confName{CVPR}
\def\confYear{2026}

\title{Are Natural-Domain Foundation Models Effective for Accelerated Cardiac MRI Reconstruction?}
\author{
Anam Hashmi$^{1}$,  Mayug Maniparambil$^{1}$,  Julia Dietlmeier$^{1}$,  Kathleen M. Curran$^{2}$,  Noel E. O'Connor$^{1}$ \and
$^{1}$Dublin City University, Ireland, $^{2}$University College Dublin, Ireland \\
{\tt\small anam.hashmi2@mail.dcu.ie, mayug.maniparambil2@mail.dcu.ie, julia.dietlmeier@insight-centre.org} \\
{\tt\small kathleen.curran@ucd.ie,  noel.oconnor@insight-centre.org}
}

\begin{document}
\maketitle
\input{sec/0_abstract}    
\input{sec/1_intro}

\input{sec/2_formatting}
\input{sec/3_finalcopy}
{
    \small
    \bibliographystyle{ieeenat_fullname}
    \bibliography{main}
}

% \clearpage

\end{document}

%% file: preamble.tex
\usepackage{multirow}

%% file: sec/0_abstract.tex
\begin{abstract}
The emergence of large-scale pretrained foundation models has transformed computer vision, enabling strong performance across diverse downstream tasks. However, their potential for physics-based inverse problems, such as accelerated cardiac MRI reconstruction, remains largely underexplored. In this work, we investigate whether natural-domain foundation models can serve as effective image priors for accelerated cardiac MRI reconstruction, and compare the performance obtained against domain-specific counterparts such as BiomedCLIP. We propose an unrolled reconstruction framework that incorporates pretrained, frozen visual encoders, such as CLIP, DINOv2, and BiomedCLIP, within each cascade to guide the reconstruction process. Through extensive experiments, we show that while task-specific state-of-the-art reconstruction models such as E2E-VarNet achieve superior performance in standard in-distribution settings, foundation-model-based approaches remain competitive. More importantly, in challenging cross-domain scenarios, where models are trained on cardiac MRI and evaluated on anatomically distinct knee and brain datasets--foundation models exhibit improved robustness, particularly under high acceleration factors and limited low-frequency sampling. We further observe that natural-image-pretrained models, such as CLIP, learn highly transferable structural representations, while domain-specific pretraining (BiomedCLIP) provides modest additional gains in more ill-posed regimes. Overall, our results suggest that pretrained foundation models offer a promising source of transferable priors, enabling improved robustness and generalization in accelerated MRI reconstruction.
Code: \href{https://github.com/Hashmi360/CMR-Recon}{github.com/Hashmi360/CMR-Recon}.
\end{abstract}

%% file: sec/1_intro.tex
\section{Introduction}
\label{sec:intro}

Cardiovascular disease continues to be the leading cause of mortality worldwide, underscoring the importance of accurate and non-invasive imaging for early diagnosis and monitoring \cite{xin2023fill}. Cardiac magnetic resonance (CMR) imaging has emerged as a powerful tool due to its ability to provide high-resolution assessment of cardiac anatomy, function, and tissue properties without the use of ionizing radiation \cite{lyu2025state, eyre2022simultaneous}.

Despite its clinical benefits, MRI acquisition is inherently slow because k-space measurements must be collected sequentially during the scan. Long acquisition times can lead to patient discomfort and motion artifacts \cite{yiasemis2022recurrent}. Accelerated MRI addresses this challenge by undersampling k-space to shorten scan duration; however, reconstructing images from such incomplete measurements is an ill-posed inverse problem (i.e. there are fewer measurements than unknowns) \cite{fabian2022humus} that requires strong prior information. 

In recent years, deep learning–based reconstruction approaches have emerged as a powerful solution to this challenge \cite{fabian2022humus, ongie2020deep}. In particular, unrolled networks have achieved state-of-the-art performance \cite{xin2023fill} by combining physics-based data consistency with learned image priors, effectively integrating the MRI acquisition model with data-driven representations. However, these conventional deep learning models are typically designed for specific tasks and often require substantial retraining or fine-tuning when deployed in new settings \cite{rajendran2025foundation}. Moreover, their clinical applicability is frequently challenged by domain shifts arising from variations in acquisition protocols, imaging settings, scanner hardware, and anatomical differences across patient populations \cite{chattopadhyay2025zero}. These limitations highlight a key challenge: how to develop reconstruction methods that generalize reliably across diverse settings.

More recently, the broader deep learning community has increasingly shifted toward large-scale foundation models, which are pretrained on extensive datasets and exhibit strong generalization across tasks and domains \cite{huix2024natural, baharoon2023evaluating}. Their success has been particularly transformative in natural language processing \cite{lewis2020bart, radford2018improving, bommasani2021opportunities}, and has recently been extended to computer vision and medical imaging \cite{radford2021learning, oquab2023dinov2, kirillov2023segment, zhang2023biomedclip, butoi2023universeg, wang2022medclip, ma2024segment}. By learning rich, transferable representations from large-scale data, these models enable strong performance even in low-data or zero-shot regimes, reducing the need for task-specific supervision. This paradigm shift suggests a promising alternative to highly specialized architectures: leveraging pretrained models as general-purpose priors.

However, applying foundation models to medical imaging, and in particular to physics-based inverse problems such as MRI reconstruction--remains largely unexplored. A key challenge lies in the domain gap between natural images used for large-scale pretraining and medical imaging data. While medical imaging is often constrained by limited annotated data due to privacy and acquisition costs \cite{huix2024natural, matsoukas2022makes, matsoukas2023pretrained}, making it well-suited for transfer learning, it is unclear whether representations learned from natural images can effectively transfer to this fundamentally different domain.

These challenges motivate a shift in perspective: rather than designing increasingly task-specific reconstruction architectures, we investigate whether the transferable representations learned by foundation models can serve as effective priors for accelerated MRI reconstruction.

Building on this perspective, we investigate whether frozen vision foundation models can serve as effective priors for physics-based MRI reconstruction, and compare natural-image-pretrained models with domain-specific counterparts such as BiomedCLIP \cite{zhang2023biomedclip}. We propose an unrolled reconstruction framework that integrates pretrained visual encoders, such as CLIP \cite{radford2021learning}, DINOv2 \cite{oquab2023dinov2}, and BiomedCLIP \cite{zhang2023biomedclip}, within each cascade of the reconstruction pipeline. The foundation models remain frozen during training and provide feature representations that guide iterative refinement, while data-consistency operations enforce fidelity to the acquired k-space measurements.

Through extensive experiments on the CMRxRecon 2023 dataset \cite{wang2023cmrxrecon}, we observe that task-specific reconstruction models, such as E2E-VarNet \cite{sriram2020end}, achieve superior performance in standard in-distribution settings, as expected due to their end-to-end optimization on domain-specific MRI data. However, a markedly different trend emerges under more challenging conditions. In cross-domain evaluations and at higher acceleration factors, corresponding to increasingly ill-posed reconstruction problems, the performance gap between the two approaches narrows substantially. In these regimes, foundation-model-based reconstructions become competitive and, in some cases, surpass task-specific baseline. We further observe that natural-image-pretrained models, such as CLIP, learn highly transferable structural representations, while domain-specific pretraining (BiomedCLIP) provides modest additional gains in more ill-posed regimes.

These findings reveal a regime-dependent trade-off: while fully supervised reconstruction network excel when sufficient domain-specific data is available, pretrained foundation models provide more robust and transferable priors under distribution shifts. Overall, this suggests that large-scale visual pretraining captures structural representations that generalize beyond natural images and can be effectively leveraged for medical image reconstruction, particularly in challenging, cross-domain settings.

In summary, our main contributions are:
\begin{itemize}
    %  \item To the best of our knowledge, we present one of the first studies investigating the use of frozen vision foundation models as priors for physics-based MRI reconstruction.
    % We introduce a novel unrolled MRI reconstruction framework that incorporates frozen vision foundation model encoders, such as CLIP, DINOv2, and BiomedCLIP, as image priors within each cascade, and provide a direct comparison between natural-image-pretrained and medical-domain-pretrained models to assess the impact of domain-specific pretraining on reconstruction performance.
    \item To the best of our knowledge, we present one of the first studies investigating frozen vision foundation models as priors for physics-based MRI reconstruction. We introduce an unrolled reconstruction framework that integrates pretrained encoders (CLIP, DINOv2, BiomedCLIP) within each cascade, and compare domain-specific (BiomedCLIP) and natural-image-pretrained models.
    
    \item We evaluate performance across both in-distribution and challenging cross-domain settings, revealing a regime-dependent behavior in which task-specific model dominate under standard conditions, while foundation-model-based approaches become increasingly competitive as reconstruction difficulty and domain shift increase.
    
    \item We show that large-scale pretrained visual representations, despite being learned from natural images, capture transferable structural priors that can generalize to MRI reconstruction across anatomically distinct domains.
\end{itemize}

\section{Background and Related Work}
\label{sec:Background and RL}

\paragraph{Inverse Problem Modeling of Accelerated MRI Reconstruction.}
Magnetic resonance imaging (MRI) acquires measurements of the underlying anatomy in the frequency domain, known as k-space. During acquisition, multiple receiver coils are used, each with a distinct spatial sensitivity profile \cite{fabian2022humus}. Let $x^\ast \in \mathbb{C}^n$ denote the target image. The measurement from the $i$-th coil is given by
\begin{equation}
k_i = \mathcal{F}(S_i x^\ast) + z_i, \quad i = 1, \dots, N,
\end{equation}
where $S_i$ denotes the coil sensitivity map, $\mathcal{F}$ the Fourier transform, and $z_i$ measurement noise. Collectively, measurements across all coils are denoted by $k = (k_1, \dots, k_N)$.

To accelerate acquisition, only a subset of k-space is sampled using an undersampling mask, yielding
\begin{equation}
\tilde{k}_i = M k_i, \quad i = 1, \dots, N,
\end{equation}
where $M$ is a binary sampling operator. The forward model can thus be written compactly as $\tilde{k} = \mathcal{A}(x^\ast)$, where $\mathcal{A}(\cdot)$ represents the undersampled multi-coil acquisition process.

Recovering $x^\ast$ from $\tilde{k}$ is an ill-posed inverse problem due to insufficient measurements. Classical approaches address this by incorporating prior knowledge, such as sparsity in a transform domain, leading to compressed sensing formulations \cite{candes2006stable, donoho2006compressed, lustig2008compressed}:
\begin{equation}
\hat{x} = \arg\min_x \; \|\mathcal{A}(x) - \tilde{k}\|^2 + \mathcal{R}(x),
\end{equation}
where $\mathcal{R}(\cdot)$ encodes prior information. In modern approaches, this prior is learned directly from data using deep neural networks, forming the basis of learning-based MRI reconstruction methods.

\paragraph{Deep Learning for Accelerated MRI Reconstruction.}
Deep learning has become the dominant paradigm for accelerated MRI reconstruction, with models learning data-driven image priors from large-scale datasets. Among these, unrolled networks achieve state-of-the-art performance by modeling reconstruction as a sequence of cascades, each corresponding to an iteration of an optimization algorithm \cite{fabian2022humus, xin2023fill, sriram2020end}. This formulation enables tight integration of MRI acquisition physics, particularly k-space undersampling, with learned image-domain priors, yielding iterative refinement schemes that progressively improve reconstruction quality. A prominent and widely adopted example is the End-to-End Variational Network (E2E-VarNet) \cite{sriram2020end}, which has established itself as a strong state-of-the-art baseline on benchmarks such as fastMRI \cite{zbontar2018fastmri,fabian2022humus}. Owing to its effectiveness and generality, E2E-VarNet has become a foundational architecture in MRI reconstruction, with many subsequent methods building upon its design.

\paragraph{Foundation Models.} Foundation models have significantly advanced transfer learning, enabling pretrained models to generalize across diverse tasks \cite{huix2024natural}. In computer vision, CLIP \cite{radford2021learning} has gained widespread attention for learning aligned image–text representations via contrastive learning, supporting strong performance across a range of downstream tasks \cite{zhao2025clip}. Recent work has explored its use in medical imaging, showing that pretrained visual features can rival or even surpass domain-specific models in certain clinical settings \cite{wang2022medclip, muller2022radiological, anand2023one}. To bridge the domain gap, BiomedCLIP \cite{zhang2023biomedclip} extends CLIP to the biomedical domain using curated image–text pairs, achieving state-of-the-art results on multiple medical benchmarks. Similarly, self-supervised vision transformers such as DINOv2 \cite{oquab2023dinov2} learn robust and transferable representations without manual annotations, achieving strong performance across diverse vision tasks \cite{song2024general, baharoon2023evaluating}. Despite these advances, existing work primarily focuses on high-level vision tasks. The role of foundation models in physics-based inverse problems, such as MRI reconstruction, remains largely unexplored.

\section{Methodology}
\label{sec:method}

The primary objective of this work is to evaluate the effectiveness of vision foundation models in the context of MRI reconstruction. In particular, we aim to assess whether features learned from large-scale pretraining, primarily on natural images--can serve as transferable priors for reconstructing undersampled MRI data.
To this end, we propose an unrolled reconstruction framework that integrates pretrained vision foundation models within a physics-based MRI reconstruction pipeline. Given undersampled multi-coil k-space measurements, the objective is to recover an image that is both consistent with the acquired data and guided by informative structural priors. The reconstruction is modeled as a sequence of cascades, each comprising a data-consistency step and a learned refinement module.

Unlike conventional approaches that learn priors solely from task-specific MRI data, our method incorporates frozen visual encoders--such as CLIP \cite{radford2021learning}, DINOv2 \cite{oquab2023dinov2}, and BiomedCLIP \cite{zhang2023biomedclip}, within each cascade. These encoders extract feature representations from intermediate reconstructions, which are used to guide iterative refinement. This design enables the model to leverage large-scale pretrained knowledge while maintaining fidelity to the underlying acquisition physics.

\paragraph{Foundation Model Encoders.}
All encoders are transformer-based models pretrained on large-scale datasets using self-supervised or multimodal objectives. We select representative models spanning both natural-image and medical-domain pretraining to study the impact of pretraining data on reconstruction performance. For all models, we use the ViT-B backbone for consistency across experiments.

\begin{itemize}
    \item \textbf{CLIP} (\textbf{C}ontrastive \textbf{L}anguage-\textbf{I}mage \textbf{P}re-training) \cite{radford2021learning} is a vision–language model trained via contrastive learning on 400 million image–text pairs \cite{radford2021learning, woerner2024navigating}, to learn aligned multimodal embeddings. This objective enables CLIP to capture rich semantic and structural features that transfer effectively across diverse visual tasks \cite{huix2024natural}. 
    % We use the OpenCLIP ViT-B model as the pretrained visual encoder.
    
    \item \textbf{DINOv2} (\textbf{DI}stillation with \textbf{NO} labels) \cite{oquab2023dinov2} is a state-of-the-art self-supervised Vision Transformer trained on a curated dataset of 142M natural images using a self-distillation objective. It learns robust and transferable visual representations without manual annotations. 
    % We use the ViT-B variant as a general-purpose feature extractor.
    
    \item \textbf{BiomedCLIP} \cite{zhang2023biomedclip} is a domain-specific extension of CLIP trained on PMC-15M, a large-scale biomedical dataset of 15M image–text pairs from PubMed Central. It captures fine-grained alignments between medical imagery and clinical context, enabling direct comparison between natural-domain and medical-domain pretrained representations. 
    % We use the ViT-B variant in our experiments.
\end{itemize}

\subsection{Network Architecture}
Our architecture follows an unrolled reconstruction paradigm, where a sequence of cascades iteratively refines the image. Each cascade consists of a data-consistency step in k-space and an image-domain refinement module. The data-consistency operation enforces agreement with acquired measurements, while the refinement module incorporates learned priors to improve reconstruction quality. In contrast to conventional approaches that rely on task-specific CNN priors, we introduce a foundation-model-guided denoiser, which leverages pretrained visual representations. This module integrates a frozen vision transformer encoder with a lightweight decoder, enabling the use of transferable features within each cascade. We first describe the architecture of the proposed foundation-model-based denoiser, followed by its integration within the unrolled reconstruction framework.

\paragraph{Foundation Model Denoiser.}
Given an intermediate reconstruction, the complex-valued MRI image is converted to a magnitude image and normalized using percentile-based scaling \cite{cekmeceli2024vision} to reduce outliers. The image is then replicated across three channels, resized to $224 \times 224$, and standardized using ImageNet \cite{deng2009imagenet} statistics to match the input distribution of pretrained encoders. The normalized image is passed through a frozen Vision Transformer encoder (CLIP, DINOv2, or BiomedCLIP), whose parameters remain fixed throughout training to ensure that reconstruction is guided by pretrained representations rather than domain-specific adaptation. We refer to the resulting models as CMR-CLIP, CMR-DINOv2, and CMR-BiomedCLIP, respectively.

Instead of relying solely on the final transformer layer, we extract intermediate features from the first six layers, which capture low-level and structural information beneficial for reconstruction \cite{huix2024natural}. These features are fused via a learnable mechanism: each layer is first aligned using LayerNorm \cite{ba2016layer}, then combined using softmax-normalized weights constrained to sum to one \cite{jiang2023clip}, enabling adaptive integration of multi-level representations. The fused patch tokens are reshaped into spatial feature maps and processed by a UNETR-style \cite{hatamizadeh2022unetr} hierarchical decoder with multi-scale skip connections from intermediate encoder layers (layers 5, 4, and 3). Each skip feature is reshaped, resized via bilinear interpolation, and projected using $1 \times 1$ convolutions before fusion.

\begin{figure}[t]
\centering

\begin{subfigure}{\linewidth}
\centering
\includegraphics[width=\linewidth]{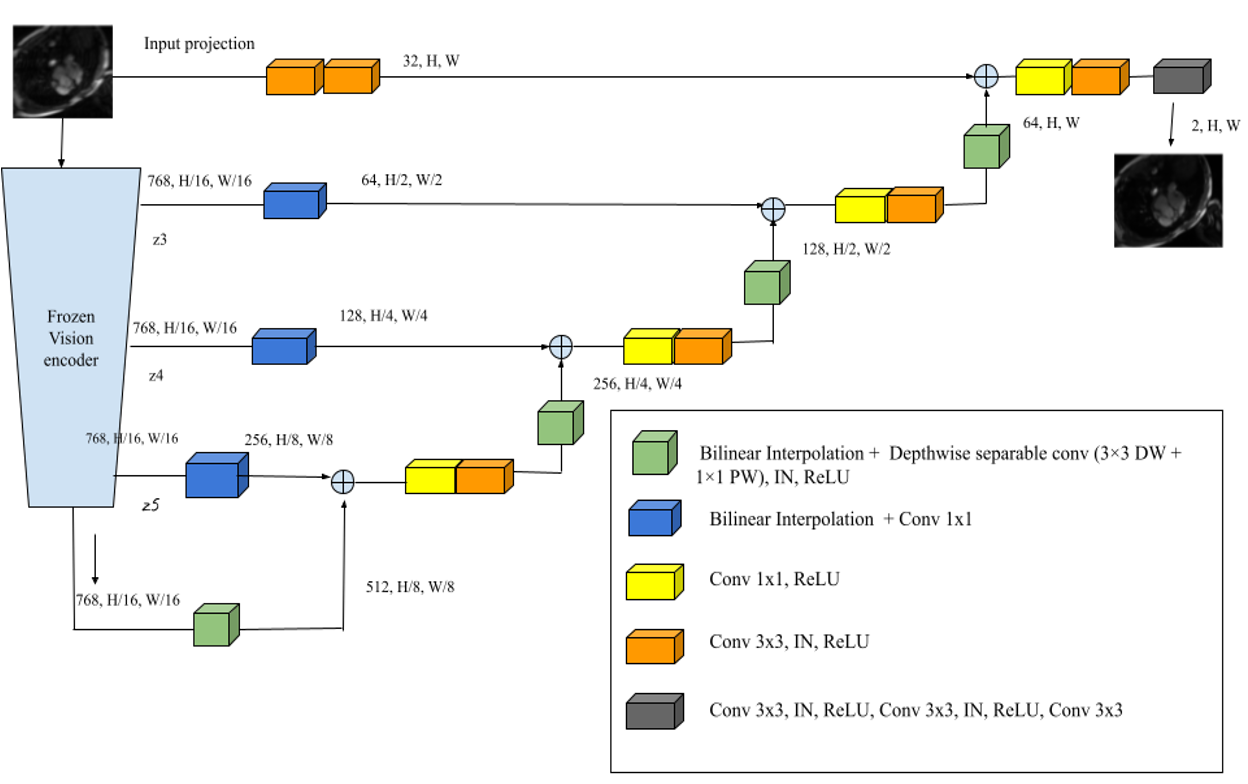}
\caption{Proposed denoiser architecture. Multi-level features from the first six layers of the frozen vision encoder are fused using LayerNorm and learnable softmax weights. Selected intermediate layers ($z_5, z_4, z_3$) provide spatially aligned skip connections, which are concatenated with decoder features at corresponding resolutions. A lightweight hierarchical decoder progressively upsamples and refines these features to generate the final reconstruction. Adapted from \cite{hatamizadeh2022unetr}}
\label{fig:denoiser}
\end{subfigure}

% \vspace{0.2em}

\begin{subfigure}{0.8\linewidth}
\centering
\includegraphics[width=\linewidth]{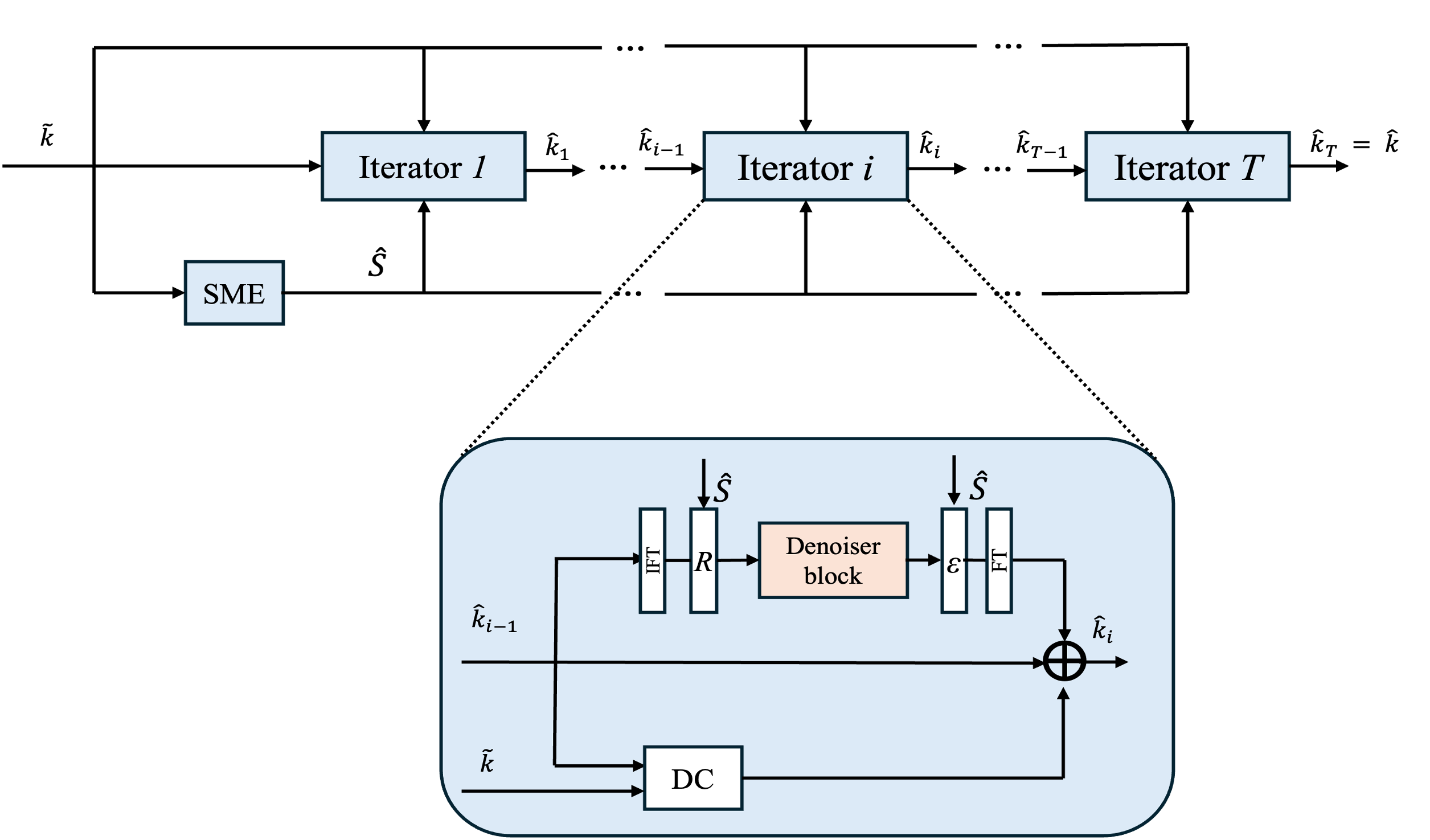}
\caption{Overview of the unrolled model architecture. Reconstruction is performed through $T$ cascades, each consisting of a data-consistency (DC) step in k-space and an image-domain refinement using the proposed denoiser. A sensitivity map estimator (SME) computes coil sensitivity maps, enabling transformations between image and multi-coil domains. Adapted from \cite{fabian2022humus}
}
\label{fig:unrolled}
\end{subfigure}

\caption{Overview of the proposed architecture.}
\label{fig:Model_architecture}
\end{figure}

Each decoding stage consists of bilinear upsampling followed by convolutional refinement. In contrast to standard UNETR \cite{hatamizadeh2022unetr} designs that employ full convolutions, we use depthwise separable convolutions (a $3 \times 3$ depthwise convolution followed by a $1 \times 1$ pointwise convolution) to improve parameter efficiency \cite{guo2019depthwise}, along with instance normalization \cite{ulyanov2016instance} and ReLU activation. At each stage, the upsampled features are concatenated with the corresponding multi-scale skip features from the encoder, as illustrated in Fig.~\ref{fig:Model_architecture}(a), and further refined using a $3 \times 3$ convolution, instance normalization, and non-linearity. This process is repeated across multiple stages, progressively recovering spatial resolution and integrating both high-level and low-level information. To preserve fine-grained details, we additionally incorporate an input-level skip connection by projecting the original complex-valued reconstruction through a shallow convolutional layer and injecting it at the final decoding stage. The final output is produced via a lightweight convolutional head, yielding a two-channel reconstruction corresponding to the real and imaginary components. Overall, the proposed denoiser combines frozen foundation model features, multi-layer transformer fusion, and a lightweight hierarchical decoder to provide transferable yet task-adapted priors for MRI reconstruction.

\subsection{ Iterative Unrolled Reconstruction}

Architectures based on unrolled optimization strategies have shown strong effectiveness in addressing inverse problems, particularly in accelerated MRI reconstruction \cite{fabian2022humus}. These methods model reconstruction as a sequence of cascades, where each stage iteratively refines the estimate by enforcing data consistency and incorporating learned image priors.

Building on the E2E-VarNet framework \cite{sriram2020end}, we adopt an unrolled reconstruction approach in the k-space domain, where the solution to the regularized inverse problem
\[
\hat{x} = \arg\min_{x} \; \left\| \mathcal{A}(x) - \tilde{k} \right\|_2^2 + \mathcal{R}(x)
\]
is approximated by unfolding the optimization into a sequence of $T$ cascades.
Each cascade corresponds to an update of the form:
\begin{equation}
\hat{k}^{t+1} = \hat{k}^{t} - \mu^t M (\hat{k}^{t} - \tilde{k}) + G(\hat{k}^{t}),
\end{equation}
where $\hat{k}^{t}$ denotes the current estimate in k-space at cascade $t$, $\mu^t$ is a learnable step size, and $G(\cdot)$ represents a learned regularization term \cite{fabian2022humus}. The second term enforces data consistency (DC) by ensuring that the reconstructed k-space remains aligned with the acquired measurements at sampled locations.

The regularization is applied in the image domain through our proposed foundation-model-guided denoiser. Specifically, the mapping $G(\cdot)$ can be expressed as:
\begin{equation}
G(k) = \mathcal{F} \big( \mathcal{E} \big( \mathcal{D} \big( \mathcal{R} (\mathcal{F}^{-1}(k)) \big) \big) \big),
\end{equation}
where $\mathcal{D}$ denotes the proposed denoiser, $\mathcal{R}(x_1, \dots, x_N) = \sum_{i=1}^{N} S_i^{*} x_i$ is the reduction operator that combines multi-coil images using the corresponding sensitivity maps, and $\mathcal{E}(x) = (S_1 x, \dots, S_N x)$ is the expansion operator that maps the combined image back to individual coil images. This formulation enables the integration of learned image-domain priors within the iterative reconstruction process while maintaining consistency with the underlying acquisition model.
We estimate coil sensitivity maps using a standard sensitivity estimation module \cite{sriram2020end}, where maps are derived from the low-frequency (ACS) region of the undersampled k-space data. These maps enable transformations between image space and multi-coil representations during reconstruction. Starting from the masked input k-space, the model applies a sequence of cascades, each performing data consistency in k-space followed by image-domain refinement using the proposed denoiser. After the final cascade, the reconstructed image is obtained via an inverse Fourier transform and combined across coils using root-sum-of-squares (RSS) \cite{sriram2020end}. This iterative formulation allows the reconstruction to be progressively refined while jointly leveraging acquisition physics and transferable pretrained visual priors.

% \begin{figure}[t]
% \centering
% \includegraphics[width=\linewidth]{sec/Denoiser_block.png}
% \caption{arch.}
% \label{fig:example}
% \end{figure}

\section{Experiments}

In this section, we present the experimental setup and evaluate the performance of the proposed method on both in-distribution and cross-domain reconstruction tasks. We first report results on the CMRxRecon \cite{wang2023cmrxrecon} dataset, where models are trained and tested on cardiac MRI data under varying acceleration factors ($\times 4$, $\times 8$, and $\times 10$). Reconstruction quality is assessed using standard metrics, including SSIM, PSNR, and NMSE. In particular, we emphasize the structural similarity index measure (SSIM) \cite{wang2003multiscale}, which is widely regarded as the primary evaluation metric in medical image reconstruction \cite{fabian2022humus}.

To further evaluate generalization, we consider challenging cross-domain settings by training models on CMRxRecon \cite{wang2023cmrxrecon} and testing them on anatomically distinct datasets from fastMRI \cite{zbontar2018fastmri}, including knee and brain MRI. This setup enables us to assess the robustness of the proposed approach under significant domain shifts.

\subsection{CMRxRecon Dataset}
The CMRxRecon dataset \cite{wang2023cmrxrecon} consists of 120 cardiac MRI cases acquired on 3T scanners, providing fully sampled dynamic cine and multi-contrast raw k-space data. The cine sequences include multiple standard cardiac views, such as short-axis (SAX) and long-axis views (2-chamber, 3-chamber, and 4-chamber), while the multi-contrast acquisitions comprise T1-weighted and T2-weighted images. For our experiments, we use the provided fully sampled data to simulate undersampling with acceleration factors of $\times 4$, $\times 8$, and $\times 10$ using uniform sampling patterns, along with corresponding masks and 24 auto-calibration (ACS) lines \cite{wang2023cmrxrecon}. The dataset is split into training, validation, and test sets using a 70\%/10\%/20\% partition, resulting in 17,916 training samples, 2,664 validation samples, and 5,112 test samples. All models are trained and evaluated on this split for in-distribution experiments.

\subsection{fastMRI Dataset}
To evaluate cross-domain generalization, we test models trained on CMRxRecon \cite{wang2023cmrxrecon} on the fastMRI dataset \cite{zbontar2018fastmri, sriram2020end}, which consists of fully sampled knee and brain MRI scans acquired on 1.5T and 3T scanners. The knee subset includes coronal proton density-weighted images with and without fat suppression, while the brain subset contains axial T1-weighted, T2-weighted, and FLAIR sequences. For evaluation, we use a subset of 1,767 knee images and 3,168 brain images. Undersampling is simulated using equispaced sampling masks with center fractions of 0.04 and 0.08 \cite{sriram2020end}, corresponding to different amounts of low-frequency information. We consider acceleration factors of $\times 4$, $\times 8$, and $\times 10$ to assess performance under varying reconstruction difficulty in cross-domain settings.

\subsection{Implementation Details}

Our framework is implemented following the standard unrolled reconstruction setting. All input images are resized to $224 \times 224$ to match the input requirements of the pretrained vision encoders. We compare our method against the state-of-the-art E2E-VarNet \cite{sriram2020end}, which we reimplement using the same architecture and hyperparameters for a fair comparison. To ensure consistency, we adopt the same sensitivity map estimation network as E2E-VarNet across all models. All reconstruction models are trained with $12$ cascades (unrolled iterations). The baseline E2E-VarNet contains approximately $29.9$M trainable parameters, while our foundation-model-based approach has approximately $56$M trainable parameters due to the additional decoder and fusion components, with the foundation model encoder kept frozen. We train separate models for acceleration factors of $\times 4$, $\times 8$, and $\times 10$. The models are optimized using the Adam optimizer with an initial learning rate of $1\times10^{-3}$, which is decayed by a factor of $0.1$ after $40$ epochs. We use no weight decay. Training is performed for $50$ epochs with early stopping based on validation performance (patience of $5$ epochs). We minimize the SSIM loss \cite{wang2003multiscale, sriram2020end} between the reconstructed and target images. For the sensitivity estimation network, we use $4$ pooling layers and $8$ base channels, consistent with prior work \cite{sriram2020end}. All experiments were conducted on NVIDIA RTX 4090 and RTX 3090 GPUs.

\section{Results}

\subsection{In-Distribution Reconstruction Results}

We first evaluate the proposed method on the CMRxRecon dataset \cite{wang2023cmrxrecon}, where models are trained and tested on cardiac MRI data under varying acceleration factors.

\begin{figure}[t]
\centering
\includegraphics[width=0.8\linewidth]{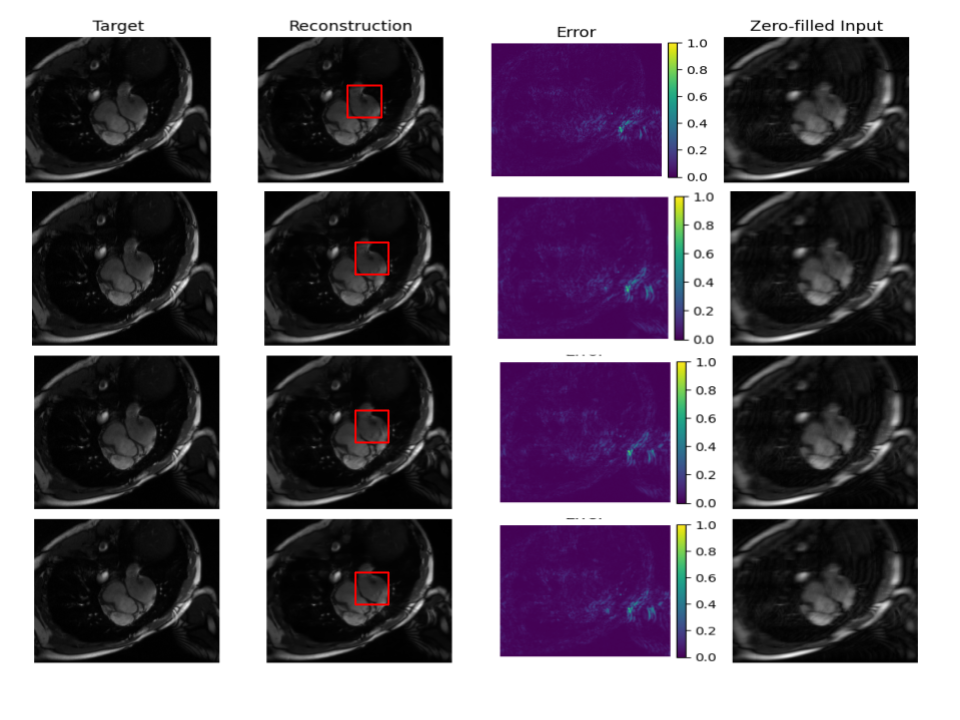}
\caption{Qualitative results at $\times 8$ acceleration. From top to bottom: E2E-VarNet, CMR-CLIP, CMR-DINOv2, and CMR-BiomedCLIP. Columns show the target image, reconstruction, error map, and zero-filled input, respectively.}
\label{fig:in_domain_results}
\end{figure}

Table~\ref{tab:cmrx_results} summarizes reconstruction performance on the CMRxRecon test set across acceleration factors of $\times 4$, $\times 8$, and $\times 10$. Qualitative comparisons are shown in Fig.~\ref{fig:in_domain_results}. As expected, the task-specific E2E-VarNet \cite{sriram2020end} consistently achieves the best results across all metrics, benefiting from end-to-end supervision tailored to MRI reconstruction.

Foundation-model-based approaches, while using frozen encoders pretrained on natural or biomedical data, remain competitive. At $\times 4$, CLIP achieves an SSIM of $0.9585$ compared to $0.9676$ for E2E-VarNet, and similar trends persist at $\times 8$ and $\times 10$, with the performance gap widening as acceleration increases. This reflects the growing difficulty of the reconstruction problem, where task-specific model retain an advantage in in-distribution settings.

Among foundation models, performance differences are relatively small. CMR-CLIP performs best at lower acceleration factors ($\times 4$ and $\times 8$), indicating that natural-image pretraining provides strong general-purpose representations. At higher acceleration ($\times 10$), CMR-BiomedCLIP becomes the strongest variant, achieving the highest SSIM ($0.9240$), suggesting that domain-specific pretraining offers advantages in more ill-posed regimes.

Overall, while task-specific architecture remain superior in standard settings, foundation models achieve competitive performance without MRI-specific training, demonstrating their ability to learn transferable structural priors.

\begin{table}[t]
\centering
\caption{Reconstruction performance on the CMRxRecon test set under different acceleration factors. Best results in bold, second-best underlined.}
\label{tab:cmrx_results}
\renewcommand{\arraystretch}{1.1}
\begin{tabular}{ccccc}
\toprule
\textbf{Acc.} & \textbf{Method} & \textbf{SSIM} $\uparrow$ & \textbf{PSNR} $\uparrow$ & \textbf{NMSE} $\downarrow$ \\
\midrule

\multirow{4}{*}{$\times 4$}
& E2E-VarNet & \textbf{0.9676} & \textbf{41.29} & \textbf{0.0154} \\
& CMR-CLIP        & \underline{0.9585} & \underline{40.12} & \underline{0.0226} \\
& CMR-DINOv2      & 0.9548 & 39.57 & 0.0249 \\
& CMR-BiomedCLIP  & 0.9557 & 39.67 & 0.0252 \\

\midrule
\multirow{4}{*}{$\times 8$}
& E2E-VarNet & \textbf{0.9502} & \textbf{38.21} & \textbf{0.0227} \\
& CMR-CLIP        & \underline{0.9359} & \underline{36.83} & \underline{0.0340} \\
& CMR-DINOv2      & 0.9340 & 36.65 & \underline{0.0340} \\
& CMR-BiomedCLIP  & 0.9358 & 36.79 & 0.0343 \\

\midrule
\multirow{4}{*}{$\times 10$}
& E2E-VarNet & \textbf{0.9417} & \textbf{37.32} & \textbf{0.0262} \\
& CMR-CLIP        & 0.9215 & 35.51 & 0.0418 \\
& CMR-DINOv2      & 0.9223 & 35.54 & 0.0419 \\
& CMR-BiomedCLIP  & \underline{0.9240} & \underline{35.73} &\underline{0.0401} \\

\bottomrule
\end{tabular}
\end{table}

\begin{figure*}[t]
\centering
\includegraphics[width=0.7\linewidth]{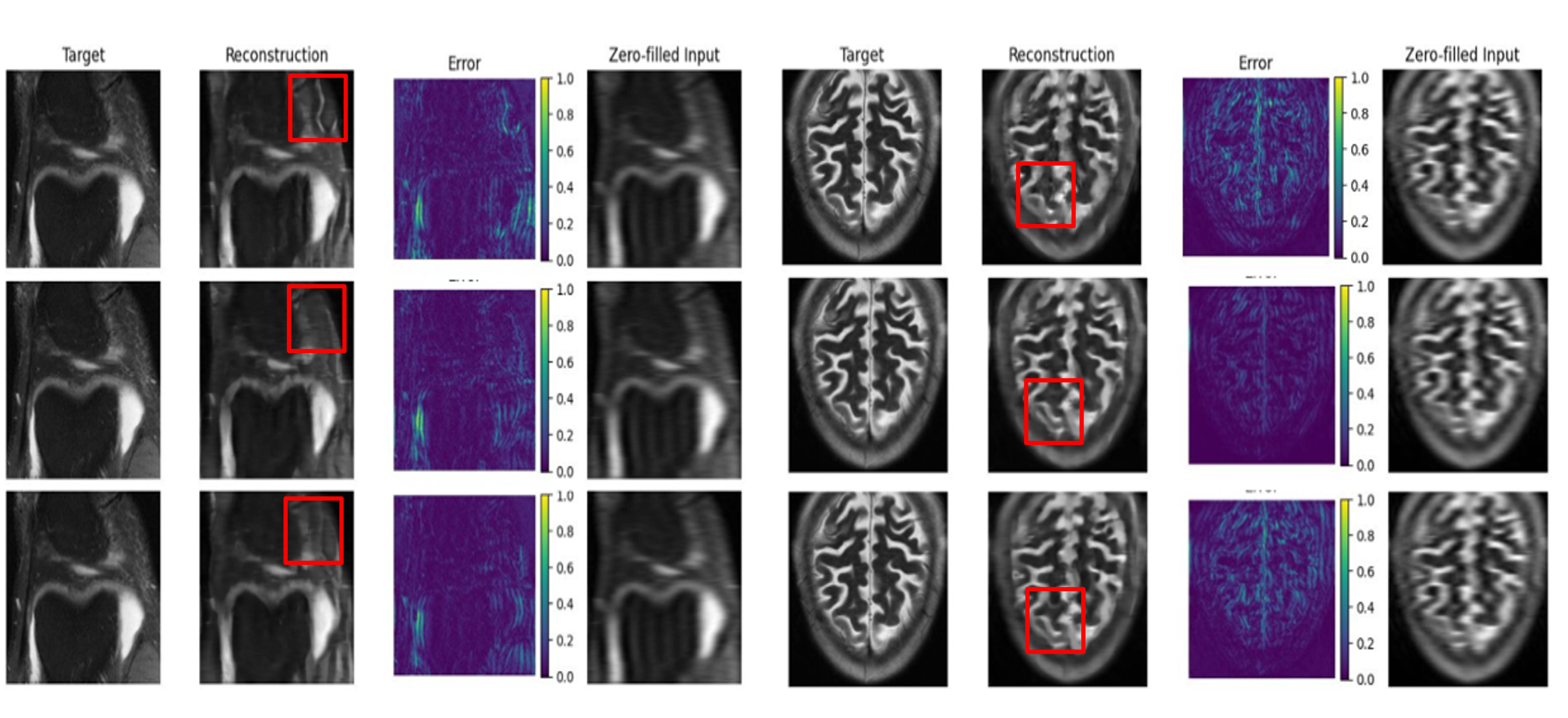}
\caption{Out-of-distribution qualitative results. Visual comparison on fastMRI knee (left) and brain (right) datasets at $\times 10$ acceleration and center fraction 0.08. From top to bottom: E2E-VarNet, CMR-CLIP, and CMR-BiomedCLIP. Columns show the target image, reconstruction, error map, and zero-filled input, respectively. Red boxes highlight regions with notable differences in structural fidelity.
}
\label{fig:ood_results_knee_brain}
\end{figure*}

\subsection{Cross-Domain Generalization Results}
We evaluate the generalization capability of the proposed approach under challenging cross-domain settings by training on the CMRxRecon dataset \cite{wang2023cmrxrecon} and testing on the fastMRI knee and brain datasets \cite{zbontar2018fastmri}. These datasets differ substantially in anatomical structure, image contrast, and acquisition characteristics, making this a stringent test of robustness. In addition to the domain shift in anatomy, we introduce a further mismatch in the sampling patterns. During training, models are exposed to equispaced undersampling masks with fixed auto-calibration signal (ACS) lines (24 central k-space lines). At test time, however, we use equispaced masks defined by center fractions of 0.04 and 0.08 \cite{sriram2020end}, where the center fraction specifies the proportion of low-frequency k-space that is fully sampled. This results in a different distribution of sampling patterns and reduces the amount of low-frequency information available, making reconstruction significantly more challenging.
Together, these shifts in anatomy, contrast, and sampling strategy create a realistic and demanding evaluation setting for assessing the robustness and transferability of learned reconstruction priors.
Quantitative results are reported in Tables~\ref{tab:ood_fastmri_knee} and \ref{tab:ood_fastmri_brain}, with corresponding qualitative comparisons shown in Fig.~\ref{fig:ood_results_knee_brain}.
% \begin{figure*}[t]
% \centering
% \includegraphics[width=0.9\linewidth]{sec/ood_results_10x_0.08.png}
% \caption{Out-of-distribution qualitative results. Visual comparison on fastMRI knee and brain datasets at $\times 10$ acceleration and center fraction 0.08, showing reconstructions from E2E-VarNet, CMR-CLIP, and CMR-BioMedCLIP, along with the zero-filled input, target image, and corresponding error maps.
% }
% \label{fig:example}
% \end{figure*}
\paragraph{Analysis and Discussion.} The cross-domain results on fastMRI knee and brain datasets reveal several important trends regarding the behavior of foundation-model-based priors under distribution shift.
% \paragraph{Performance under mild vs. severe acceleration.}
At lower acceleration ($\times4$), E2E-VarNet consistently achieves the best performance across both knee and brain datasets. This is expected, as fully supervised reconstruction models are optimized for the training distribution and can effectively exploit dataset-specific priors when sufficient measurements are available. However, as reconstruction becomes more challenging, through higher acceleration ($\times 8$, $\times 10$) and reduced center fraction (0.04), a clear shift in behavior emerges. The performance of E2E-VarNet degrades more rapidly, while foundation-model-based approaches exhibit more stable performance and consistently match or outperform E2E-VarNet.
This trend is consistent across both anatomical domains. On the knee dataset, CMR-BiomedCLIP achieves the best performance at $\times 8$, while CMR-CLIP and CMR-BiomedCLIP outperform E2E-VarNet at $\times 10$, particularly under reduced center fraction. Similarly, on the brain dataset, foundation models surpass E2E-VarNet at $\times 8$ and $\times 10$, with CMR-BiomedCLIP achieving the strongest overall performance at the highest acceleration. These gains are most pronounced under center fraction 0.04, where limited low-frequency information increases reconstruction ambiguity and places greater reliance on learned priors.

Comparing foundation models, CMR-CLIP performs strongly at moderate acceleration, indicating that natural-image pretraining captures highly transferable structural representations. CMR-BiomedCLIP provides additional gains in more ill-posed regimes, suggesting that domain-specific pretraining becomes beneficial as reconstruction difficulty increases. In contrast, CMR-DINOv2 consistently underperforms across settings, particularly under severe domain shift, likely due to the absence of cross-modal alignment in its pretraining objective \cite{liu2025data}, which may limit its ability to encode higher-level structural priors necessary for robust MRI reconstruction.

Overall, these results suggest a regime-dependent behavior: task-specific  model excel in in-distribution and well-conditioned settings, while pretrained foundation models provide more robust and transferable priors under severe undersampling and domain shift. This highlights the potential of leveraging large-scale pretrained representations as a complementary alternative to conventional task-specific reconstruction pipelines, particularly in scenarios where training data is mismatched.

\begin{table*}[t]
\centering
\caption{OOD reconstruction performance on the FastMRI Knee dataset under different acceleration factors. Best results in bold, second-best underlined.}
\label{tab:ood_fastmri_knee}
\renewcommand{\arraystretch}{1.1}
\begin{tabular}{c c ccc ccc}
\toprule
\textbf{Acc} & \textbf{Method} 
& \multicolumn{3}{c}{\textbf{Center frac = 0.08}} 
& \multicolumn{3}{c}{\textbf{Center frac = 0.04}} \\
\cmidrule(lr){3-5} \cmidrule(lr){6-8}
& 
& SSIM $\uparrow$ & PSNR $\uparrow$ & NMSE $\downarrow$
& SSIM $\uparrow$ & PSNR $\uparrow$ & NMSE $\downarrow$ \\
\midrule

\multirow{4}{*}{4x}
& E2E-VarNet  & \textbf{0.8061} & \textbf{31.06} & \textbf{0.0280} & \textbf{0.7769} & \textbf{29.21} & \textbf{0.0412} \\
& CMR-CLIP        & \underline{0.7864} & \underline{30.88} & \underline{0.0288} & \underline{0.7387} &\underline{27.92} & \underline{0.0548} \\
& CMR-DINOv2      & 0.7412 & 27.50 & 0.0647 & 0.6873 & 21.91 & 0.2781 \\
& CMR-BiomedCLIP  & 0.7848 & 30.66 & 0.0302 & 0.7310 & 27.60 & 0.0591 \\

\midrule
\multirow{4}{*}{8x}
& E2E-VarNet  & 0.7161 & 28.59 & 0.0479 & 0.6552 & 25.88 & 0.0876 \\
& CMR-CLIP        & \underline{0.7232} & \underline{28.72} & \underline{0.0470} & \underline{0.6668} & \underline{26.17} & \underline{0.0820} \\
& CMR-DINOv2      & 0.7137 & 27.33 & 0.0646 & 0.6566 & 25.61 & 0.0932 \\
& CMR-BiomedCLIP  & \textbf{0.7285} & \textbf{28.84} & \textbf{0.0457} & \textbf{0.6712} & \textbf{26.26} & \textbf{0.0804} \\

\midrule
\multirow{4}{*}{10x}
& E2E-VarNet  & 0.6956 & 28.08 & 0.0542 & 0.6287 & 25.26 & 0.1014 \\
& CMR-CLIP        & \textbf{0.7150} & \underline{28.36} & \underline{0.0509} & \textbf{0.6576} & \textbf{26.08} & \textbf{0.0839} \\
& CMR-DINOv2      & 0.6960 & 27.03 & 0.0684  & 0.6244 & 24.73 & 0.1148 \\
& CMR-BiomedCLIP  & \underline{0.7133} & \textbf{28.55} & \textbf{0.0491} & \underline{0.6545} & \underline{25.91} & \underline{0.0873} \\

\bottomrule
\end{tabular}
\end{table*}

\begin{table*}[t]
\centering
\caption{OOD reconstruction performance on the FastMRI Brain dataset under different acceleration factors. Best results in bold, second-best underlined.}
\label{tab:ood_fastmri_brain}
\renewcommand{\arraystretch}{1.1}
\begin{tabular}{c c ccc ccc}
\toprule
\textbf{Acc} & \textbf{Method} 
& \multicolumn{3}{c}{\textbf{Center frac = 0.08}} 
& \multicolumn{3}{c}{\textbf{Center frac = 0.04}} \\
\cmidrule(lr){3-5} \cmidrule(lr){6-8}
& 
& SSIM $\uparrow$ & PSNR $\uparrow$ & NMSE $\downarrow$
& SSIM $\uparrow$ & PSNR $\uparrow$ & NMSE $\downarrow$ \\
\midrule

\multirow{4}{*}{4x}
& E2E-VarNet  & \textbf{0.8209} & \textbf{30.29} & \textbf{0.0241} & \textbf{0.7670} & \textbf{27.23} & \textbf{0.0483} \\
& CMR-CLIP        & 0.7962 & \underline{29.69} & \underline{0.0281}  & \underline{0.7208} & \underline{25.79} & \underline{0.0682} \\
& CMR-DINOv2      & 0.7321 & 25.65 & 0.0808  & 0.6531 & 20.59 & 0.3796 \\
& CMR-BiomedCLIP  & \underline{0.7985} & 29.52 & 0.0290  & 0.7163 & 25.43 & 0.0747 \\

\midrule
\multirow{4}{*}{8x}
& E2E-VarNet  & 0.7363 & \underline{27.25} & \underline{0.0473} & 0.6416 & 23.74 & 0.1069 \\
& CMR-CLIP        & \underline{0.7391} & \textbf{27.36} & \textbf{0.0462} & \underline{0.6505} & \textbf{23.99} & \textbf{0.1006} \\
& CMR-DINOv2      & 0.7203 & 25.28 & 0.0753 & 0.6300 & 22.86 & 0.1309 \\
& CMR-BiomedCLIP  & \textbf{0.7447} & 27.14 & 0.0483 & \textbf{0.6514} & \underline{23.88} & \underline{0.1035} \\

\midrule
\multirow{4}{*}{10x}
& E2E-VarNet  & 0.7234 & \underline{26.84} & \underline{0.0520} & 0.6209 & \underline{23.33} & 0.1172 \\
& CMR-CLIP        & \underline{0.7261} & 26.22 & 0.0594 & \underline{0.6312} & 23.32 & \underline{0.1166} \\
& CMR-DINOv2      & 0.7142 & 25.42 & 0.0730 & 0.6135 & 22.63 & 0.1407 \\
& CMR-BiomedCLIP  & \textbf{0.7326} & \textbf{26.87} & \textbf{0.0515} & \textbf{0.6374} & \textbf{23.47} & \textbf{0.1134} \\

\bottomrule
\end{tabular}
\end{table*}

%\begin{figure*}[t]
%\centering
%\includegraphics[width=0.9\linewidth]%{sec/ood_results_10x_0.08.png}
%\caption{Visualization of reconstruction results.}
%\label{fig:example}
%\end{figure*}

\section{Conclusion}

We present the first study investigating frozen vision foundation models as priors for physics-based MRI reconstruction, introducing an unrolled framework that integrates pretrained visual encoders within each cascade. While task-specific models such as E2E-VarNet achieve superior performance in standard in-distribution settings, foundation-model-based approaches demonstrate greater robustness under cross-domain shifts and severe undersampling, where the reconstruction problem becomes increasingly ill-posed. Notably, as reconstruction difficulty increases, these models provide more stable and transferable priors, narrowing and in some cases surpassing the performance gap with the supervised baseline. Furthermore, we find that natural-image-pretrained models such as CLIP already capture highly transferable structural representations, with domain-specific pretraining BiomedCLIP offering modest additional gains in challenging regimes. Overall, our findings highlight the potential of large-scale pretrained visual representations as a complementary source of priors for improving robustness and generalization in MRI reconstruction and, more broadly, in physics-based inverse problems.

\section*{Acknowledgements}
This work was supported by Taighde \'Eireann–Research Ireland under Grant numbers 18/CRT/6183 \& 12/RC/2289 P2.

%% file: main.bib
@String(AAAI = {AAAI})

@inproceedings{xin2023fill,
  title={Fill the k-space and refine the image: Prompting for dynamic and multi-contrast MRI reconstruction},
  author={Xin, Bingyu and Ye, Meng and Axel, Leon and Metaxas, Dimitris N},
  booktitle={International Workshop on Statistical Atlases and Computational Models of the Heart},
  pages={261--273},
  year={2023},
  organization={Springer}
}

@article{lyu2025state,
  title={The state-of-the-art in cardiac mri reconstruction: Results of the cmrxrecon challenge in miccai 2023},
  author={Lyu, Jun and Qin, Chen and Wang, Shuo and Wang, Fanwen and Li, Yan and Wang, Zi and Guo, Kunyuan and Ouyang, Cheng and T{\"a}nzer, Michael and Liu, Meng and others},
  journal={Medical Image Analysis},
  volume={101},
  pages={103485},
  year={2025},
  publisher={Elsevier}
}

@article{eyre2022simultaneous,
  title={Simultaneous multi-parametric acquisition and reconstruction techniques in cardiac magnetic resonance imaging: basic concepts and status of clinical development},
  author={Eyre, Katerina and Lindsay, Katherine and Razzaq, Saad and Chetrit, Michael and Friedrich, Matthias},
  journal={Frontiers in cardiovascular medicine},
  volume={9},
  pages={953823},
  year={2022},
  publisher={Frontiers Media SA}
}

@inproceedings{yiasemis2022recurrent,
  title={Recurrent variational network: a deep learning inverse problem solver applied to the task of accelerated MRI reconstruction},
  author={Yiasemis, George and Sonke, Jan-Jakob and S{\'a}nchez, Clarisa and Teuwen, Jonas},
  booktitle={Proceedings of the IEEE/CVF conference on computer vision and pattern recognition},
  pages={732--741},
  year={2022}
}

@article{fabian2022humus,
  title={Humus-net: Hybrid unrolled multi-scale network architecture for accelerated mri reconstruction},
  author={Fabian, Zalan and Tinaz, Berk and Soltanolkotabi, Mahdi},
  journal={Advances in Neural Information Processing Systems},
  volume={35},
  pages={25306--25319},
  year={2022}
}

@article{ongie2020deep,
  title={Deep learning techniques for inverse problems in imaging},
  author={Ongie, Gregory and Jalal, Ajil and Metzler, Christopher A and Baraniuk, Richard G and Dimakis, Alexandros G and Willett, Rebecca},
  journal={IEEE Journal on Selected Areas in Information Theory},
  volume={1},
  number={1},
  pages={39--56},
  year={2020},
  publisher={IEEE}
}

@misc{huix2024natural,
  title={Are Natural Domain Foundation Models Useful for Medical Image Classification? In 2024 IEEE/CVF Winter Conference on Applications of Computer Vision (WACV). Waikoloa, HI, USA: IEEE;[cited 2024 Aug 27]. 7619--7628},
  author={Huix, JP and Ganeshan, AR and Haslum, JF and S{\"o}derberg, M and Matsoukas, C and Smith, K},
  year={2024}
}

@article{rajendran2025foundation,
  title={Foundation models in medical image analysis: A systematic review and meta-analysis},
  author={Rajendran, Praveenbalaji and Safari, Mojtaba and He, Wenfeng and Hu, Mingzhe and Wang, Shansong and Zhou, Jun and Yang, Xiaofeng},
  journal={arXiv preprint arXiv:2510.16973},
  year={2025}
}

@article{baharoon2023evaluating,
  title={Evaluating general purpose vision foundation models for medical image analysis: An experimental study of dinov2 on radiology benchmarks},
  author={Baharoon, Mohammed and Qureshi, Waseem and Ouyang, Jiahong and Xu, Yanwu and Aljouie, Abdulrhman and Peng, Wei},
  journal={arXiv preprint arXiv:2312.02366},
  year={2023}
}

@inproceedings{lewis2020bart,
  title={BART: Denoising sequence-to-sequence pre-training for natural language generation, translation, and comprehension},
  author={Lewis, Mike and Liu, Yinhan and Goyal, Naman and Ghazvininejad, Marjan and Mohamed, Abdelrahman and Levy, Omer and Stoyanov, Veselin and Zettlemoyer, Luke},
  booktitle={Proceedings of the 58th annual meeting of the association for computational linguistics},
  pages={7871--7880},
  year={2020}
}

@article{radford2018improving,
  title={Improving language understanding by generative pre-training},
  author={Radford, Alec and Narasimhan, Karthik and Salimans, Tim and Sutskever, Ilya and others},
  year={2018},
  publisher={San Francisco, CA, USA}
}

@article{chattopadhyay2025zero,
  title={Zero-shot domain generalization of foundational models for 3d medical image segmentation: an experimental study},
  author={Chattopadhyay, Soumitri and Demir, Basar and Niethammer, Marc},
  journal={arXiv preprint arXiv:2503.22862},
  year={2025}
}

@article{bommasani2021opportunities,
  title={On the opportunities and risks of foundation models},
  author={Bommasani, Rishi and Hudson, Drew A and Adeli, Ehsan and Altman, Russ and Arora, Simran and von Arx, Sydney and Bernstein, Michael S and Bohg, Jeannette and Bosselut, Antoine and Brunskill, Emma and others},
  journal={arXiv preprint arXiv:2108.07258},
  year={2021}
}

@inproceedings{radford2021learning,
  title={Learning transferable visual models from natural language supervision},
  author={Radford, Alec and Kim, Jong Wook and Hallacy, Chris and Ramesh, Aditya and Goh, Gabriel and Agarwal, Sandhini and Sastry, Girish and Askell, Amanda and Mishkin, Pamela and Clark, Jack and others},
  booktitle={International conference on machine learning},
  pages={8748--8763},
  year={2021},
  organization={PmLR}
}

@article{oquab2023dinov2,
  title={Dinov2: Learning robust visual features without supervision},
  author={Oquab, Maxime and Darcet, Timoth{\'e}e and Moutakanni, Th{\'e}o and Vo, Huy and Szafraniec, Marc and Khalidov, Vasil and Fernandez, Pierre and Haziza, Daniel and Massa, Francisco and El-Nouby, Alaaeldin and others},
  journal={arXiv preprint arXiv:2304.07193},
  year={2023}
}

@inproceedings{kirillov2023segment,
  title={Segment anything},
  author={Kirillov, Alexander and Mintun, Eric and Ravi, Nikhila and Mao, Hanzi and Rolland, Chloe and Gustafson, Laura and Xiao, Tete and Whitehead, Spencer and Berg, Alexander C and Lo, Wan-Yen and others},
  booktitle={Proceedings of the IEEE/CVF international conference on computer vision},
  pages={4015--4026},
  year={2023}
}

@inproceedings{wang2022medclip,
  title={Medclip: Contrastive learning from unpaired medical images and text},
  author={Wang, Zifeng and Wu, Zhenbang and Agarwal, Dinesh and Sun, Jimeng},
  booktitle={Proceedings of the 2022 Conference on Empirical Methods in Natural Language Processing},
  pages={3876--3887},
  year={2022}
}

@article{ma2024segment,
  title={Segment anything in medical images},
  author={Ma, Jun and He, Yuting and Li, Feifei and Han, Lin and You, Chenyu and Wang, Bo},
  journal={Nature communications},
  volume={15},
  number={1},
  pages={654},
  year={2024},
  publisher={Nature Publishing Group UK London}
}

@inproceedings{butoi2023universeg,
  title={Universeg: Universal medical image segmentation},
  author={Butoi, Victor Ion and Ortiz, Jose Javier Gonzalez and Ma, Tianyu and Sabuncu, Mert R and Guttag, John and Dalca, Adrian V},
  booktitle={Proceedings of the IEEE/CVF International Conference on Computer Vision},
  pages={21438--21451},
  year={2023}
}

@article{zhang2023biomedclip,
  title={Biomedclip: a multimodal biomedical foundation model pretrained from fifteen million scientific image-text pairs},
  author={Zhang, Sheng and Xu, Yanbo and Usuyama, Naoto and Xu, Hanwen and Bagga, Jaspreet and Tinn, Robert and Preston, Sam and Rao, Rajesh and Wei, Mu and Valluri, Naveen and others},
  journal={arXiv preprint arXiv:2303.00915},
  year={2023}
}

@article{matsoukas2023pretrained,
  title={Pretrained vits yield versatile representations for medical images},
  author={Matsoukas, Christos and Haslum, Johan Fredin and Sorkhei, Moein and S{\"o}derberg, Magnus and Smith, Kevin},
  journal={arXiv preprint arXiv:2303.07034},
  year={2023}
}

@inproceedings{matsoukas2022makes,
  title={What makes transfer learning work for medical images: Feature reuse \& other factors},
  author={Matsoukas, Christos and Haslum, Johan Fredin and Sorkhei, Moein and S{\"o}derberg, Magnus and Smith, Kevin},
  booktitle={Proceedings of the IEEE/CVF conference on computer vision and pattern recognition},
  pages={9225--9234},
  year={2022}
}

@article{candes2006stable,
  title={Stable signal recovery from incomplete and inaccurate measurements},
  author={Candes, Emmanuel J and Romberg, Justin K and Tao, Terence},
  journal={Communications on Pure and Applied Mathematics: A Journal Issued by the Courant Institute of Mathematical Sciences},
  volume={59},
  number={8},
  pages={1207--1223},
  year={2006},
  publisher={Wiley Online Library}
}

@article{donoho2006compressed,
  title={Compressed sensing},
  author={Donoho, David L},
  journal={IEEE Transactions on information theory},
  volume={52},
  number={4},
  pages={1289--1306},
  year={2006},
  publisher={IEEE}
}

@article{lustig2008compressed,
  title={Compressed sensing MRI},
  author={Lustig, Michael and Donoho, David L and Santos, Juan M and Pauly, John M},
  journal={IEEE signal processing magazine},
  volume={25},
  number={2},
  pages={72--82},
  year={2008},
  publisher={IEEE}
}

@article{jiang2023clip,
  title={From clip to dino: Visual encoders shout in multi-modal large language models},
  author={Jiang, Dongsheng and Liu, Yuchen and Liu, Songlin and Zhao, Jin'e and Zhang, Hao and Gao, Zhen and Zhang, Xiaopeng and Li, Jin and Xiong, Hongkai},
  journal={arXiv preprint arXiv:2310.08825},
  year={2023}
}

@inproceedings{sriram2020end,
  title={End-to-end variational networks for accelerated MRI reconstruction},
  author={Sriram, Anuroop and Zbontar, Jure and Murrell, Tullie and Defazio, Aaron and Zitnick, C Lawrence and Yakubova, Nafissa and Knoll, Florian and Johnson, Patricia},
  booktitle={International conference on medical image computing and computer-assisted intervention},
  pages={64--73},
  year={2020},
  organization={Springer}
}

@article{wang2023cmrxrecon,
  title={CMRxRecon: an open cardiac MRI dataset for the competition of accelerated image reconstruction},
  author={Wang, Chengyan and Lyu, Jun and Wang, Shuo and Qin, Chen and Guo, Kunyuan and Zhang, Xinyu and Yu, Xiaotong and Li, Yan and Wang, Fanwen and Jin, Jianhua and others},
  journal={arXiv preprint arXiv:2309.10836},
  year={2023}
}

@article{zbontar2018fastmri,
  title={fastMRI: An open dataset and benchmarks for accelerated MRI},
  author={Zbontar, Jure and Knoll, Florian and Sriram, Anuroop and Murrell, Tullie and Huang, Zhengnan and Muckley, Matthew J and Defazio, Aaron and Stern, Ruben and Johnson, Patricia and Bruno, Mary and others},
  journal={arXiv preprint arXiv:1811.08839},
  year={2018}
}

@article{zhao2025clip,
  title={CLIP in medical imaging: A survey},
  author={Zhao, Zihao and Liu, Yuxiao and Wu, Han and Wang, Mei and Li, Yonghao and Wang, Sheng and Teng, Lin and Liu, Disheng and Cui, Zhiming and Wang, Qian and others},
  journal={Medical Image Analysis},
  volume={102},
  pages={103551},
  year={2025},
  publisher={Elsevier}
}

@inproceedings{muller2022radiological,
  title={Radiological reports improve pre-training for localized imaging tasks on chest x-rays},
  author={M{\"u}ller, Philip and Kaissis, Georgios and Zou, Congyu and Rueckert, Daniel},
  booktitle={International Conference on Medical Image Computing and Computer-Assisted Intervention},
  pages={647--657},
  year={2022},
  organization={Springer}
}

@article{anand2023one,
  title={One-shot localization and segmentation of medical images with foundation models},
  author={Anand, Deepa and Singhal, Vanika and Shanbhag, Dattesh D and KS, Shriram and Patil, Uday and Bhushan, Chitresh and Manickam, Kavitha and Gui, Dawei and Mullick, Rakesh and Gopal, Avinash and others},
  journal={arXiv preprint arXiv:2310.18642},
  year={2023}
}

@article{song2024general,
  title={General purpose image encoder dinov2 for medical image registration},
  author={Song, Xinrui and Xu, Xuanang and Yan, Pingkun},
  journal={arXiv preprint arXiv:2402.15687},
  year={2024}
}

@inproceedings{cekmeceli2024vision,
  title={Do vision foundation models enhance domain generalization in medical image segmentation?},
  author={Cekmeceli, Kerem and Himmetoglu, Meva and Tombak, Guney I and Susmelj, Anna and Erdil, Ertunc and Konukoglu, Ender},
  booktitle={European Conference on Computer Vision},
  pages={185--200},
  year={2024},
  organization={Springer}
}

@inproceedings{deng2009imagenet,
  title={Imagenet: A large-scale hierarchical image database},
  author={Deng, Jia and Dong, Wei and Socher, Richard and Li, Li-Jia and Li, Kai and Fei-Fei, Li},
  booktitle={2009 IEEE conference on computer vision and pattern recognition},
  pages={248--255},
  year={2009},
  organization={Ieee}
}

@article{ba2016layer,
  title={Layer normalization},
  author={Ba, Jimmy Lei and Kiros, Jamie Ryan and Hinton, Geoffrey E},
  journal={arXiv preprint arXiv:1607.06450},
  year={2016}
}

@inproceedings{hatamizadeh2022unetr,
  title={Unetr: Transformers for 3d medical image segmentation},
  author={Hatamizadeh, Ali and Tang, Yucheng and Nath, Vishwesh and Yang, Dong and Myronenko, Andriy and Landman, Bennett and Roth, Holger R and Xu, Daguang},
  booktitle={Proceedings of the IEEE/CVF winter conference on applications of computer vision},
  pages={574--584},
  year={2022}
}

@article{ulyanov2016instance,
  title={Instance normalization: The missing ingredient for fast stylization},
  author={Ulyanov, Dmitry and Vedaldi, Andrea and Lempitsky, Victor},
  journal={arXiv preprint arXiv:1607.08022},
  year={2016}
}

@inproceedings{guo2019depthwise,
  title={Depthwise convolution is all you need for learning multiple visual domains},
  author={Guo, Yunhui and Li, Yandong and Wang, Liqiang and Rosing, Tajana},
  booktitle={Proceedings of the AAAI conference on artificial intelligence},
  volume={33},
  number={01},
  pages={8368--8375},
  year={2019}
}

@inproceedings{wang2003multiscale,
  title={Multiscale structural similarity for image quality assessment},
  author={Wang, Zhou and Simoncelli, Eero P and Bovik, Alan C},
  booktitle={The thrity-seventh asilomar conference on signals, systems \& computers, 2003},
  volume={2},
  pages={1398--1402},
  year={2003},
  organization={Ieee}
}

@article{liu2025data,
  title={Data or Language Supervision: What Makes CLIP Better than DINO?},
  author={Liu, Yiming and Zhang, Yuhui and Ghosh, Dhruba and Schmidt, Ludwig and Yeung-Levy, Serena},
  journal={arXiv preprint arXiv:2510.11835},
  year={2025}
}

@inproceedings{woerner2024navigating,
  title={Navigating data scarcity using foundation models: A benchmark of few-shot and zero-shot learning approaches in medical imaging},
  author={Woerner, Stefano and Baumgartner, Christian F},
  booktitle={International Workshop on Foundation Models for General Medical AI},
  pages={30--39},
  year={2024},
  organization={Springer}
}
